\documentclass[journal]{IEEEtran}

\IEEEoverridecommandlockouts 

\usepackage{xcolor} 
\usepackage{amsthm}

\usepackage{amsmath,amssymb,amsthm}

\newcounter{cAss}
\newcounter{cAssSaved}
\newcommand{\Ass}[1]{\ensuremath{\boldsymbol{ h}_{\mathbf{#1}}}}

\newlength\asswidth
\newenvironment{assumptions}{%
  \begin{list}{\Ass{\arabic{cAss}}:}{%
    \setcounter{cAssSaved}{\value{cAss}}%
    \usecounter{cAss}%
    \setcounter{cAss}{\value{cAssSaved}}%
    \setlength\topsep{.5ex plus .2ex minus .2ex}%
    \setlength\itemsep{.5ex}%
    \settowidth\asswidth{\Ass{\arabic{cAss}}:}%
    \addtolength\asswidth{\labelsep}%
    \setlength\leftmargin{\asswidth}%
    \setlength\labelwidth{\asswidth}%
  }%
}{\end{list}}

\makeatletter
\newcommand{\@endgadget}[1]{%
  {\unskip\nobreak\hfil\penalty50\hskip1em\hbox{}\nobreak\hfil#1%
  \parfillskip=0pt\finalhyphendemerits=0\par}%
}

\newcommand{\@Endofsymbol}{$\triangledown$}   

\newcommand{\Endofremark}{\@endgadget{\@Endofsymbol}}
\makeatother

\newcommand{\beq}{\begin{equation}}
\newcommand{\eeq}{\end{equation}}

\usepackage{enumitem}
\usepackage{balance}

\newcommand{\bag}{\begin{aligned}}
\newcommand{\eag}{\end{aligned}}
\usepackage{float}
\newcommand{\sign}{\operatorname{sign}}
\usepackage{amssymb}
\usepackage{graphicx}

\usepackage{amsmath, amssymb}
\usepackage{mathrsfs}
\usepackage{balance}

\newcommand{\cA}{\mathcal{A}}

\newcommand{\cL}{\mathcal{L}}

\newtheorem{theorem}{Theorem}

\newtheorem{definition}{Definition}

\newtheorem{remark}{Remark}

\usepackage{color}
\usepackage{bm}
\usepackage{balance}

\usepackage{url} 

\begin{document}
\title{On Finite- and Fixed-Time Stabilization of Abstract  Nonlinear Systems with Well-Posedness Guarantees
}
\author{Kamal Fenza, Moussa Labbadi and Mohamed Ouzahra
\thanks{K. Fenza and M. Ouzahra are with the  Laboratory of  Mathematics and Applications to Engineering Sciences,  Sidi Mohamed Ben Abdellah University, Fes, Morocco (e-mail: \textsl{kamal.fenza@usmba.ac.ma}, \textsl{mohamed.ouzahra@usmba.ac.ma})}
\thanks{M. Labbadi is with the Aix-Marseille University, LIS UMR CNRS 7020, 13013 Marseille, France (e-mail: \textsl{moussa.labbadi@lis-lab.fr}). 
}
}

\maketitle

\begin{abstract}
This paper addresses the problem of stabilization for infinite-dimensional systems. In particular, we design nonlinear stabilizers for both linear and nonlinear abstract systems. We focus on two classes of systems: the first class comprises linear abstract systems subject to matched perturbations, while the second class encompasses fully nonlinear abstract systems. Our main objective is to synthesize state-feedback controllers that guarantee finite- or fixed-time stability of the closed-loop system, along with possible estimation of the settling time. For the first class, the presence of persistent perturbations introduces significant challenges in the well-posedness analysis, particularly due to the discontinuous nature of the control law. To address this, we employ maximal monotone operator theory to rigorously establish the existence and uniqueness of solutions, extending classical results from continuous abstract systems. For the second class, which includes nonlinearities, we further show that the proposed feedback law ensures fixed-time stability and well-posedness of the closed-loop system, again using maximal monotone theory. The results provide a unified framework for robust, finite /fixed-time stabilization in the presence of discontinuities and nonlinearities in infinite-dimensional settings.
\end{abstract}

\section{Introduction}
Control of partial differential equations (PDEs) has attracted considerable attention in recent decades, not only because of the rich mathematical challenges it poses but also due to its broad spectrum of applications in engineering, physics, biology, and other scientific domains. Designing effective control strategies for systems governed by PDEs is of paramount importance for addressing real-world problems such as fluid dynamics, heat conduction, chemical processes, population dynamics, and wave propagation.  

Among the diverse classes of PDE systems, abstract systems occupy a central place. The stabilization of such systems has been extensively studied in the literature for several decades (see, for instance, \cite{ma2025integral}), where particular emphasis was placed on the problem of exponential stability. In more recent years, however, finite-time stabilization has emerged as a compelling alternative, offering stronger convergence guarantees by ensuring that the system reaches equilibrium in a bounded time interval. The origins of finite-time stabilization can be traced back to early works on optimal control, where the so-called Fuller phenomenon was first observed~\cite{Fuller1960}.  

In classical mechanics, numerous examples illustrate how finite-dimensional systems naturally exhibit finite-time stabilization. A notable case arises when dry friction acts as the dominant dissipative force, driving the system to its equilibrium state in finite time. Motivated by such phenomena, Ha\"im Br\'ezis proposed a conjecture concerning the finite-time stabilization of a system given by  
\begin{equation}\label{eq:1.4}
    u_{tt} - \Delta u + \mu_c \, \sign(u_t) + \mu_v u_t \ni  0,
\end{equation}  
where $\sign: \mathbb{R} \to \mathcal{P}(\mathbb{R})$ denotes the set-valued sign function, defined by
\(
\sign(x) =  
1\)  if $x > 0$,
$-1$ if $x < 0$,
$[-1,1]$ if $x = 0$.
In this model, the term $\mu_c \, \sign(u_t)$ accounts for \emph{Coulomb (dry) friction}, while $\mu_v u_t$ represents a possible \emph{viscous friction} component. Br\'ezis conjectured that, at least in the case $\mu_v = 0$, the system’s equilibrium position is reached within a finite time. Furthermore, in the case where $\mu_{v} > 0$, it was shown in~\cite{baji2007asymptotics} that 
equation~\eqref{eq:1.4} admits solutions which converge exponentially fast toward their limit, 
as well as solutions which stabilize within a finite time.  Inspired by Bogolioubov and Mitropolski \cite{Brogliato1999,pohjolainen1982robust}, this conjecture also appears in Haraux's thesis \cite{Haraux1978Thesis}, which motivated the works of Cabannes and Bamberger--Cabannes. For a one-dimensional domain ($\Omega = (0,1)$), Eq.~(\ref{eq:1.4}) models a vibrating string with friction. Cabannes \cite{Cabannes1978, Cabannes1981} obtained partial finite-time stabilization results for specific initial data, while the general case remains open.

Finite-time stabilization results for homogeneous linear and quasilinear hyperbolic systems are discussed in \cite{CoronNguyen2022,CoronNguyen2020}. 
In the presence of such as perturbations in $L^{\infty}$, the problem is interesting almost of researchers in PDEs and variables structure systems include so-called sliding modes~\cite{Utkin1992} or unit control~\cite{orlov2002discontinuous, cortes2006finite,coron2017null,efimov2021finite,coron2021boundary,nguyen2024rapid,polyakov2016finite,han2024input, 10, 15, 17}  extended the notion of homogeneity to infinite-dimensional systems. They developed a linear stabilizing feedback and a Lyapunov function for homogeneous dilation to obtain finite-time stability results, which were then applied to the wave and heat equations.
Additionally, the author in \cite{ouzahra2021finite} proved finite-time stability for the bilinear reaction-diffusion equation at a settling time that depends on the initial state (in the absence of disturbances).  For abstract bilinear and linear systems without matched perturbation, the problem of finite-time stabilization via the explicit construction of a Lyapunov function and decomposition method  has been investigated in~\cite{fenza2025decomposition}. In that work, the theoretical results are complemented by applications to both parabolic and hyperbolic partial differential equations.

A powerful tool for achieving
finite-time stabilization and rejecting matched perturbations is sliding mode control (SMC), particularly in the control of dynamical systems \cite{sogore2020, sob, balogoun2023, pilloni2024sliding,pisano2012boundary,orlov2004robust,pisano2011tracking,orlov2013boundary,orlov1995sliding,chitour2024sliding, Levaggi2002a, Levaggi2002b}. It has been extensively studied in the context of infinite-dimensional systems, where representation and state-space methods provide a solid theoretical foundation for analysis and controller design \cite{bensoussan2007, curtain2020}. In hyperbolic and parabolic PDE control, SMC strategies, including super-twisting algorithms, have proven effective in ensuring robustness and finite-time convergence despite system uncertainties \cite{balogoun2023,coron2021, guo2012, guo2013}.  These approaches  make use of  Lyapunov-based techniques to construct strict stability proofs, as demonstrated in the development of super-twisting controllers \cite{moreno2012}.
Furthermore, boundary sliding mode control methods offer robust solutions for handling disturbances in heat and wave equations, where conventional control techniques have difficulty \cite{guo2012,pisano2012,  wang2015}. For instance, second-order sliding mode controllers have been successfully applied to heat processes with unbounded perturbations, ensuring strong disturbance rejection capabilities \cite{pisano2012}. Similarly, the combination of SMC with active disturbance rejection control has led to significant improvements in the stabilization of wave and Euler-Bernoulli beam equations under boundary disturbances \cite{guo2012, guo2013}. The application of SMC to cascaded partial differential equations--ordinary differential equations systems has further demonstrated its flexibility in managing complex multi-scale dynamics, providing efficient stabilization mechanisms even under severe input constraints \cite{wang2015}. By integrating SMC with these advanced techniques, researchers continue to expand its applicability to various classes of PDEs, ensuring both theoretical rigor and practical feasibility in real-world control systems.

Inspired by the book of Brézis~\cite{brezis1973operateurs} on well-posedness and by our conference paper presented at CDC~\cite{fenza2025finite}, we first establish that, for a linear abstract problem subject to matched perturbations, the closed-loop system achieves finite-time stability under a discontinuous control law. Moreover, we show that the system is maximally monotone in the sense of Brézis’ definition. We then extend the analysis to fixed-time stability, where the settling time is shown to be independent of the initial conditions. In this case, both maximal monotonicity and fixed-time stability of the closed-loop system are rigorously proved.
The second contribution of this paper concerns the study of a class of nonlinear abstract systems. Specifically, we propose a nonlinear feedback law that guarantees fixed-time stability of the closed-loop system. The existence of solutions is established using the theory of maximal monotone operators. Furthermore, explicit estimates of the settling time and a complete stability analysis are provided.
Finally, the theoretical developments are illustrated through applications to the heat equation, considered under both matched perturbations and nonlinearities. These examples demonstrate the applicability and effectiveness of the proposed approach.

This article represents an extended journal version of our earlier conference contribution~\cite{fenza2025finite}. The main advancements of the present work can be summarized as follows:  
\begin{enumerate}
    \item \textbf{Generalization of the problem setting.} Whereas the conference paper focused primarily on finite-time stability of the reaction--diffusion equation with matched perturbations, together with abstract nonlinear systems of the form~(\ref{eq:system_nonlinear}), the current article broadens the framework to address both finite-time and fixed-time stabilization of a wider class of linear and nonlinear abstract systems.  

    \item \textbf{Strengthened theoretical contributions.} We provide refined and more detailed  proofs concerning the existence, uniqueness, and stability of solutions. These enhancements establish a stronger theoretical foundation for the obtained stabilization results.  
\end{enumerate}

The remainder of this paper is organized as follows. In Section~\ref{sec:Problem}, we formulate the problem and state the key assumptions underlying our analysis. Section~\ref{sec:Monotone} is dedicated to the foundational results, including a review of maximal monotone operators and evolution equations. In Section~\ref{sec:mainresults}, we present the main contributions of this work, focusing on finite- and fixed-time stabilization as well as well-posedness. The proofs of these results are provided in Section~\ref{sec:Proofs}. Section~\ref{sec:applications} demonstrates the applicability of the abstract framework through the example of a heat equation. Finally, Section~\ref{sec:conclusions} offers concluding remarks and outlines directions for future research.

\emph{Notations.} 

\begin{itemize}
    \item $\mathbb{R}$ denotes the field of real numbers, with $\mathbb{R}^+ = [0, +\infty)$,  $\mathbb{R}^{-} = (-\infty,0]$  {\color{black} and \(\mathbb{R}^* = \mathbb{R} \setminus \{0\}\) denotes the set of nonzero real numbers.}
    \item $\mathcal{H}$ is a real Hilbert space endowed with the inner product $\langle \cdot, \cdot \rangle$ and its corresponding norm $\|\cdot\|$. 
    \item Let $C(\mathbb{R}^{+}, \mathcal{H})$ denote the space of all continuous functions defined on $\mathbb{R}^+$ and taking values in $\mathcal{H}$.
    \item $I$ represents the identity operator on $\mathcal{H}$. 
    \item $B^{*}$ is the adjoint of the operator $B$.
    \item Let \( A: D(A) \subset \mathcal{H} \to \mathcal{H} \) be a (possibly) unbounded operator, where \( D(A) \) denotes its domain.
    \item $\mathcal{H}^p(\Omega, \mathbb{R})$ denotes the Sobolev space of functions $\Omega \to \mathbb{R}$, where $\Omega \subset \mathbb{R}^n$ is an open connected set with a smooth boundary.
    \item $L^\infty(\Omega)$ is the space of essentially bounded, Bochner-measurable functions with the norm $\|\eta\|_\infty := \operatorname{ess\,sup}_{s \in \Omega} \|\eta(s)\|$.
    \item $C_c^\infty(\Omega)$ is the set of infinitely smooth functions $\mathbb{R}^n \to \mathbb{R}$ with compact support in $\Omega$.
    \item $\mathcal{H}_0^p(\Omega, \mathbb{R})$ is the closure of $C_c^\infty(\Omega)$ in the norm of $\mathcal{H}^p(\Omega, \mathbb{R})$.
   \item We denote by $D(\cA)$ its domain, by $\cA^\ast$ its adjoint.
   \item We denote by $L^{1}_{\mathrm{loc}}\big(0,+\infty;\mathcal{H}\big)$
the space of all measurable functions 
\(h : (0,+\infty) \to \mathcal{H}\) 
such that, for every finite interval \([0,T]\) with \(T>0\), one has 
\[
\int_{0}^{T} \|h(t)\|_{\mathcal{H}} \, dt < +\infty.
\]
That is, \(h\) is integrable on every bounded subinterval of \((0,+\infty)\).
\end{itemize}
 \section{Problem Statement}\label{sec:Problem}
We consider the nonlinear control system
\begin{equation}\label{eq:system_nonlinear}
\Sigma_1:\;
\begin{cases}
\dot y(t) = A y(t) + f(y(t))\,y(t) + B u(y(t)), & t>0, \\[2pt]
y(0) = y_0,
\end{cases}
\end{equation}
where $y(\cdot)$ denotes the state, $u$ is a $\mathcal{U}$-valued control input with $\mathcal{U}$ a real Hilbert space, and the state space $\mathcal{H}$ is also a Hilbert space. The linear operator $A: D(A) \rightarrow \mathcal{H}$ is unbounded and generates  a quasi-contractive $C_0$ semigroup $S(t)$ of type $\omega \in \mathbb{R}$ on $\mathcal{H}$ (in particular $\forall\, \mathrm{t} \geq 0:\; \|S(\mathrm{t})\| \leq \exp({\omega \mathrm{t}})$ and $\langle Ay, y\rangle \leq \omega \|y \|^2$, for all $y\in \mathcal{H}$), and $B: \mathcal{U} \rightarrow \mathcal{H}$ is bounded and linear (i.e. $B\in \cL(\mathcal{U} ,\mathcal{H} )$).  The nonlinear term $f: \mathcal{H} \rightarrow \mathbb{R}$ is scalar-valued, and we define the operator $F: \mathcal{H} \rightarrow \mathcal{H}$ by
\[
F(y) = -f(y) y.
\]
Before analyzing the full nonlinear dynamics of \eqref{eq:system_nonlinear}, we focus on  the finite /fixed -time stabilization of the following  perturbed  abstract control system:
\begin{equation}\label{eq:Abstract}
\Sigma_2:\;
\begin{cases}
\dot y(t) = A y(t) + B u(y(t))+ \eta(t), & t>0, \\[2pt]
y(0) = y_0,
\end{cases}
\end{equation}
where  $\eta$ is a the matched perturbation  from $[0, +\infty)$ into $\mathcal{H}$.

The finite-time stabilization of a particular instance of system~\eqref{eq:Abstract}, corresponding to the reaction–diffusion equation, has recently been investigated in~\cite{fenza2025finite}, where the analysis was carried out using the maximum principle. It is necessary to extend this approach to general abstract systems and to consider the fixed-time stability property. In doing so, the well-posedness of the closed-loop system must also be studied.

In the second part of this paper, we focus on the finite- and fixed-time stabilization of system~\eqref{eq:system_nonlinear}. The goal is to design a feedback control and determine a finite settling time $T$ such that all trajectories of~\eqref{eq:system_nonlinear} converge to zero within this time and remain at zero thereafter. In addition to finite-time stability, we consider fixed-time stability, where the settling time is independent of the initial conditions. Since the closed-loop system is nonlinear, it is necessary to rigorously study the well-posedness of the solutions.

We make the following assumptions:

\begin{assumptions}
\item The map $F$ can be decomposed as $F = F_{1} + F_{2}$, where $F_{1}$ is maximal monotone on $\mathcal{H}$ and $F_{2}$ is Lipschitz on $\mathcal{H}$:
\[
\|F_2(s_1) - F_2(s_2)\| \leq \kappa \|s_1 - s_2\|, \quad \forall s_1, s_2 \in \mathcal{H},
\]
for some $\kappa > 0$.\label{ass:nonlinear}
\end{assumptions}

\begin{assumptions}
\item There exists $\beta > 0$ such that 
\[
\|B^{*}y\|_{\mathcal{U}} \geq \beta \|y\|, \quad \forall y \in \mathcal{H}.
\]\label{ass:coercivity}
\end{assumptions}
\begin{assumptions}
   \item There exists a constant \(M>0\) such that, for all \(t \geq 0\), one has
\[
\|\eta(t)\|\leq M.
\]
\end{assumptions}
\begin{remark}
$\bullet$ Assumption \Ass{2} guarantees the coercivity of the operator $BB^{*}$, a standard requirement in infinite-dimensional control theory to achieve global finite-time stabilization (see, e.g., \cite{sogore2020,sob}).  \\
$\bullet$ Under Assumption \Ass{2}, the operator $BB^{*}$ is bounded and invertible from $\mathcal{H}$ onto $\mathcal{H}$.  \\
$\bullet$  Assumption \Ass{3} is equivalent to 
$\eta \in L^{\infty}(0,+\infty;\mathcal{H})$,  which is a condition that is fundamental for analyzing the finite- and fixed-time stability of the considered systems. Moreover. Assumption \Ass{3} guarantees that the matched perturbation $\eta$ belongs to  $L^{1}(0,T;\mathcal{H})$ or every  $T>0$, ensuring its integrability over any finite time interval, which is crucial for establishing the well-posedness of the systems.
\Endofremark
\end{remark}
The system \eqref{eq:system_nonlinear} with nonlinearities is well-studied using the theory of evolution semigroups \cite{pazy2012semigroups,engel2000one}. Recent work \cite{polyakov2021input, polyakov2024finite} addressed input-to-state stability of locally Lipschitz infinite-dimensional systems using homogeneous techniques.

In this paper, the precise definitions of finite-time and fixed-time stability are not restated for brevity. For a rigorous treatment of finite-time stability, the reader is referred to~\cite{bhat2000finite,bhat2005geometric}, while the concept of fixed-time stability is thoroughly presented in~\cite{polyakov2011nonlinear}.
\section{ Review on maximal monotone operator and  evolution equations}\label{sec:Monotone}
This section is devoted to some essential properties of maximal monotone operators and  their  relation with  existence and uniqueness theory for  evolution equations.
 \begin{definition} ([\cite{brezis1973operateurs}, pp. 20-22]):  The operator $\mathcal{A}: D(\mathcal{A}) \subset \mathcal{H} \rightarrow \mathcal{H}$ is said to be monotone if  $\langle \mathcal{A}(x)-\mathcal{A}(y), x-y\rangle \geq 0$, for all $x, y \in D(\mathcal{A})$ (i.e. $-\mathcal{A}$ is dissipative). If in addition,   the graph of $\mathcal{A} $  is not properly contained in the graph of any multivalued   monotone operator in $\mathcal{H}$, we say that  $\mathcal{A} $ is maximal monotone.
\end{definition} 
The subdifferential of a lower semicontinuous convex function provides a fundamental example of a maximal monotone operator. This establishes a strong connection between the theory of nonlinear  maximal monotone operators and convex analysis. Based on this fact, we can state the following result:
\begin{theorem}([\cite{brezis1973operateurs}, p. 25])\label{C1}
Let $h$ be a proper, lower-semicontinuous convex function on $\mathcal{H}$. Then, the subdifferential of $h$ is maximal monotone in $\mathcal{H}$.
\end{theorem}
It is well known that the sum of two monotone operators is itself monotone.  However, this property does not extend in general to the case of maximal monotone operators. 
This issue is of significant interest due to its applications in the study of partial differential equations.  In this context, we present the following results:
 \begin{theorem} ([\cite{brezis1973operateurs}, p. 34]):\label{C2}
Let $\mathcal{A}$ be a maximal monotone operator on $\mathcal{H}$ and $\mathcal{B}$ a Lipschitzian monotone operator from $\mathcal{H}$ into $\mathcal{H}$. Then the operator $\mathcal{A}+\mathcal{B}$ is maximal monotone.
\end{theorem} 
\begin{theorem}  ([\cite{brezis1973operateurs}, p. 36]): \label{C3}
Let $\mathcal{A}$ and $\mathcal{B}$ be two maximal monotone operators on $\mathcal{H}$. If $D(\mathcal{A}) \cap \operatorname{Int}(D(\mathcal{B})) \neq \emptyset$, where $\operatorname{int}(D(\mathcal{B}))$ denotes the interior of $D(\mathcal{B})$, then $\mathcal{A}+\mathcal{B}$ is a maximal monotone operator.
\end{theorem} 
Now, we introduce the notions of weak and strong  solutions of the following evolution equation:
\begin{eqnarray}\label{equation}
	\begin{cases}
		\frac{d}{d t}  x(t)+\mathcal{A} x(t)\ni v(t) , \quad    \hspace*{0.1cm} t>0,\\
		x(0)=x_{0}\in \mathcal{H}.  \\
	\end{cases}
\end{eqnarray}
where $\mathcal{A}: D(\mathcal{A}) \subset \mathcal{H} \rightarrow \mathcal{H}$ is a  (possibly nonlinear) operator and $v$ is a given function from $[0, +\infty)$ into $\mathcal{H}$.
\\
\begin{definition} ([\cite{brezis1973operateurs}, p. 64]):
 Let  $v\in L^{1}_{\mathrm{loc}}\big(0,+\infty;\mathcal{H}\big)$.  The continuous  function $x:  [0, +\infty)\rightarrow \mathcal{H}$ 
 is called a strong  solution of  (\ref{equation}) if  it satisfies \\
$\bullet$ $x$ is absolutely continuous on any compact subset of   $(0, +\infty)$,\\
$\bullet$ $x(t) \in D(\mathcal{A})$ and $\frac{d x}{d t}(t)+\mathcal{A} x(t)\ni f(t)$, a.e. for all  $t\in (0, +\infty)$.
\end{definition} 
\begin{definition} ([\cite{brezis1973operateurs}, p. 64]):  A continuous  function $x:  [0, +\infty)\rightarrow \mathcal{H}$ is a weak solution for  (\ref{equation}) if there exist sequences $(v_{n})\in L^{1}_{\mathrm{loc}}\big(0,+\infty;\mathcal{H}\big)$ such that $v_{n}\rightarrow v$ in $ L^{1}_{\mathrm{loc}}\big(0,+\infty;\mathcal{H}\big)$ 
and $(x_{n})\in C([0, +\infty), \mathcal{H})$ such that  $x_{n}$ is a strong
solution  of $\frac{d }{d t} x_{n}(t)+\mathcal{A} x_{n}(t)\ni v_{n}(t) $ and $x_{n}\rightarrow x$ uniformly on $[0, T]$, for any $T > 0$.
\end{definition} 
The next results establish the existence and uniqueness of the weak solution.
\begin{theorem}  ([\cite{brezis1973operateurs}, p.  65]):\label{C4}
Let  $\mathcal{A}: D(\mathcal{A}) \subset \mathcal{H} \rightarrow \mathcal{H}$ be a  maximal monotone operator and $v\in L^{1}_{\mathrm{loc}}\big(0,+\infty;\mathcal{H}\big)$. Then, for all $x_{0}\in \overline{D(\mathcal{A})}$  the  system  (\ref{equation})  admits  a unique   weak solution  $x \in C([0,+\infty), \mathcal{H})$.
\end{theorem} 
\begin{theorem}([\cite{brezis1973operateurs}, p.  105])\label{C5}
Let $A$ be a maximally monotone operator, $x_{0} \in \overline{D(\mathcal{A})}$, $\alpha  > 0$ and $v \in L^{1}_{\mathrm{loc}}\big(0,+\infty;\mathcal{H}\big)$, with $T > 0$.  Then there exists a unique  weak solution $x \in C([0,+\infty), \mathcal{H})$ to the following evolution equation:
\begin{eqnarray}\label{equation1}
	\begin{cases}
		\frac{d}{d t}  x(t) +\mathcal{A} x(t) - \alpha ~ x(t)\ni v(t) , \quad    \hspace*{0.1cm} t>0,\\
		x(0)=x_{0}\in \mathcal{H},  \\
	\end{cases}
\end{eqnarray}
\end{theorem}
\section{Main Results}\label{sec:mainresults}
The main results of our paper are presented in this section in terms of finite-/fixed-time stability and well-posedness.
\subsection{ Finite/fixed-Time Stabilization of Abstract  Systems $\Sigma_2$}
We consider the abstract linear system~\eqref{eq:Abstract}, for which the  well-posedness in closed loop will be investigated. Before turning to the analysis of existence and uniqueness of solutions, we first discuss the choice of a feedback law ensuring finite- and fixed-time stabilization. We begin with the study of finite-time stability.  Our approach is based on the construction of a suitable Lyapunov function candidate $V$, 
followed by the derivation of sufficient conditions on the system parameters that guarantee  the differential inequality~\cite{bhat2000finite}.
\begin{equation}\label{ineq:Lyap}
\dot{V} + c V^{\theta} \leq 0, 
\qquad \theta \in (0,1), \; c>0.
\end{equation}
To motivate the feedback design and the choice of the Lyapunov function, 
let us formally compute the time derivative of the ``energy'' associated with the state:
\[
\frac{1}{2}\,\frac{d}{dt}\|y(t)\|^{2}
= \langle Ay(t), y(t) \rangle 
+ \langle Bu(t), y(t) \rangle 
+ \langle \eta(t), y(t) \rangle .
\]
In particular, when the operator $A$ is dissipative and the disturbance is absent, that is $\eta(t)=0$ for all $t \geq 0$, dissipativity can be enforced by selecting the feedback law
\[
u(t) = - \|B^{*}y(t)\|^{\zeta} B^{*}y(t),
\]
for a suitable real exponent $\zeta$.  Following related studies and specific examples in the literature  (see \cite{17,ouzahra2021finite}), a natural and effective choice is obtained by setting  $\beta = -\mu$ with $0<\mu<1$, which achieves finite-time stabilization 
of~\eqref{eq:Abstract}.\\
On the other hand, for a quasi-contractive semigroup $S(t)$ of type $\omega$, in the presence of uncertainties or external disturbances $\eta$, a more robust control strategy is required. 
In this case, we propose the following feedback control law:
\begin{equation}\label{control}
u(y(t)) = - \lambda B^{*}y(t) - \rho\, \operatorname{sign}(B^{*}y(t)),
\end{equation}
where  $\rho >0$ is the gain control,  \(\lambda \in \mathbb{R}^{+}\) is  an appropriate coefficient and will be determined in the stability analysis.
 and the set-valued map is defined as:
\begin{equation}
	\operatorname{sign} (y) =\begin{cases}   \{\frac{y}{\|y\|_{\mathcal{U}}}\} & \text { if}~  y ~ \neq 0, \\ \mathcal{B}_{\mathcal{U}} & \text { if } y=0.\end{cases}
\end{equation}
where $\mathcal{B}_{\mathcal{U}}$   denotes the closed unit ball in $\mathcal{U}$. \\
Noting that when the type of the semigroup satisfies $\omega \leq 0$, 
the semigroup $(S(t))_{t \geq 0}$ is contractive. 
In this situation, the linear component of the feedback law is no longer required, and one can simply set $\lambda = 0$, as  in~\cite{17,ouzahra2021finite}.
Therefore, without loss of generality, we restrict our attention to the case  $\omega \geq 0$, where the presence of the additional linear component in the control 
becomes essential for fixed and finite-time  stabilization.\\
For our purpose, \textcolor{black}{combining {\color{black}$\Sigma_2$} and {\color{black}(\ref{control})} leads to the following differential inclusion}:
\begin{eqnarray}\label{KAMAKkbbb}
	\begin{cases}
		\frac{d }{d t}y(t) +\Gamma (y(t))-\omega~  y(t) \ni \eta(t), ~~ &   t\in\mathbb{R}^{+}, \\
	
		y(0=y_{0},
	\end{cases}
\end{eqnarray} 
with  $\Gamma: D(A) \rightarrow P(\mathcal{H})$ is  defined by $\Gamma(y)=-Ay+ \omega~ y + \lambda~  BB^{*}y+ \rho B~\operatorname{sign}(B^{*}y),$ for all $y\in D(A)$.\\
In the following result  we give a representation of the solution of the differential inclusion system {\color{black} (\ref{KAMAKkbbb})} that is based on nonlinear semigroup theory.
\begin{theorem}[Well-posedness]\label{thm:M1}
Let $A$ generate a quasi-contractive $C_0$-semigroup of type $\omega \geq 0$ and assume that \Ass{3} holds. Then, for all $y_{0}\in \mathcal{H}$,  the  system {\color{black} (\ref{eq:Abstract})} has a unique global  weak solution satisfies that $y\in C([0,+\infty), \mathcal{H})$.
\end{theorem} 
Let $T>0$ and let $y:[0,T)\to\mathcal{H}$ be the weak solution of \eqref{KAMAKkbbb}. 
We define
\[
\mathcal{K}(t):=-\lambda B^{*}y(t)-\rho B(r(t))+\eta(t),
\]
 where $r(t)\in \operatorname{sign}(B^{*}y(t))$. Since $y$ is continuous (as a weak solution)  and the mapping  $y \mapsto B^{*}y$ is  bounded, it follows that 
$\mathcal{K}\in L^{1}(0,T;\mathcal{H})$ for every $T>0$. 
Consequently, form \cite{pazy2012semigroups}, the solution $y$ admits the variation-of-constants representation
\[
y(t)=S(t)y_{0}+\int_{0}^{t} S(t-s)\,\mathcal{K}(s)\,ds, 
\qquad t\in[0,T).
\]
Moreover, if $y_{0}\in D(A)$, then the solution $y$ is strong: 
$y(t)\in D(A)$ for all $t\in[0,+\infty)$, the map $t\mapsto y(t)$ is continuously differentiable 
on $(0,T)$, and \eqref{KAMAKkbbb} is satisfied almost everywhere on the interval $(0,T)$.\\
\\
We now state our main result concerning the global finite-time stabilization of (\ref{KAMAKkbbb}) toward the origin equilibrium,  then let us consider the following assumption:
 \begin{theorem}[Global finite-time stability]\label{thm:M2} Assume    that \Ass{2} and \Ass{3} hold. Then,  the system  (\ref{eq:Abstract}) is globally finite-time stable under  the following control:
\begin{equation}\label{con}
u(y(t))  =\begin{cases}   \{ - \frac{\omega}{\beta^2}  B^{*}y(t)- \frac{\rho B^{*}y(t)}{\left\|B^{*}y(t)\right\|_{\mathcal{U}}}\} & \text { if}~  y(t) ~ \neq 0, \\  \{ - \frac{\omega}{\beta^2}  B^{*}y(t)\}+\rho\mathcal{B}_{\mathcal{U}} & \text { if } y(t)=0.\end{cases}
\end{equation}
where  that the gain control is such that 
\begin{equation}\label{gain}
    \rho >\max\left\{\frac{M}{\beta},\; M\sqrt{\bigl\|(BB^{*})^{-1}\bigr\|}\right\}.
\end{equation}
Furthermore, the settling time admits the estimate  $T(y_0)\leq\frac{\left\|y_{0}\right\|}{  \rho \beta -M}$.
\end{theorem}
\begin{remark}
The choice of the gain control $\rho$ with Assumption~\Ass{3} provides a sufficient condition to ensure that the origin is an equilibrium point of system~(\ref{eq:Abstract}). 
More precisely, under \Ass{3}, the system admits the trivial solution \(y(t)\equiv 0\), which remains invariant under the dynamics. 
This observation highlights that the origin can be regarded as a natural equilibrium candidate, forming the basis for a subsequent analysis of its stability properties.
\Endofremark
\end{remark}
We reconsider system (\ref{eq:Abstract}) under the same assumptions on the operators 
$A$ and $B$ as stated in Subsection~II. In the preceding analysis, we  established that system (\ref{eq:Abstract})  possesses the property of finite-time stability; however, the associated settling-time function was shown to depend explicitly on the initial condition, which may lead to variability in the convergence rate. In the present section, we aim to strengthen this result by proposing a suitably designed feedback control law that not only ensures convergence in finite time but also provides a uniform upper bound for the settling-time function, independent of the initial state. This stronger stabilization property  is referred to as \textit{fixed-time stability}.  \\
We adopt the same methodology as in the finite-time stability analysis 
and construct a feedback control law such that the Lyapunov function 
associated with the closed-loop system satisfies the following 
differential inequality~\cite{polyakov2011nonlinear}.
\begin{equation}\label{fix-time}
\dot{V} + c_{1} V^{\theta} + c_{2} V^{\alpha} \leq 0,
\end{equation}
where   $\theta \in (0,1)$, $\alpha > 1$,  $c_{1}, c_{2} > 0$.
\begin{theorem}[Global fixed-time stability]\label{thm:M3}
    Assume that the Assumptions \Ass{2} and \Ass{3}  hold. Then,  the system  (\ref{eq:Abstract})  is globally fixed-time stable under the following control:
\begin{equation}\label{con2}
u(y(t))  =\begin{cases}   \{ - \frac{\omega}{\beta^2}  B^{*}y(t)- \frac{\rho B^{*}y(t)}{\left\|B^{*}y(t)\right\|_{\mathcal{U}}}-\left\|B^{*}y(t)\right\|_{\mathcal{U}}^{\zeta}B^{*}y(t)\},\\ \quad \quad \quad\quad \text { if}~  y(t) ~ \neq 0, \\ \rho \mathcal{B}_{\mathcal{U}}, \quad  \text { if } y(t)=0.\end{cases}
\end{equation}
where  $\zeta >0$, $\omega$  denotes  the type of the semigroup $S(t)$, and the control gain $\rho$ satisfies (\ref{gain}).\\
The settling time admits the  following estimate:
\begin{equation}\label{T}
    T(y_0)\leq \frac{1}{ \rho \beta -M }+ \frac{1}{ \zeta \beta^{\zeta+ 2}}.
\end{equation}

\end{theorem}
\subsection{Finite/fixed-time stability results for abstract nonlinear system $\Sigma_1$}
In this section, we investigate nonlinear  system $\Sigma_1$, with respect to control design, aiming to achieve fixed-time stability while ensuring well-posedness of the corresponding closed-loop systems.
\subsubsection{Control design}
We consider system~($\Sigma_1$) and aim to study its well-posedness in closed-loop.  Before addressing the existence and uniqueness of solutions,  we first discuss the choice of a feedback law to achieve fixed-time stabilization.  
Our approach relies on constructing a suitable Lyapunov function candidate $V$,  and then providing sufficient conditions on the system parameters to guarantee the  differential inequality (\ref{fix-time}).


To gain insight into the feedback law, we consider the Lyapunov function candidate $V(y) = \|y\|^2$ 
and formally compute its time derivative:
\begin{align}
\frac{1}{2}\,\frac{\mathrm{d}}{\mathrm{dt}}\|y(t)\|^2
&= \langle A y(t),\, y(t)\rangle  + \langle B u(t),\, y(t)\rangle \notag\\
& + \langle f(y(t))\,y(t),\, y(t)\rangle.
\end{align}
Thus, for instance when $A$ is dissipative and $f=0$, the dissipativity of the system can be ensured by selecting the control as $u(t)=-\|B^{*}y\|^{\zeta} B^{*}y$ for a suitable real exponent $\zeta$. Based on related studies and specific examples in the literature (see~\cite{17,ouzahra2021finite}), 
a natural and effective choice for achieving global finite-time stabilization is  $\zeta = -\mu$, where $0 < \mu < 1$.\\
Now, let us assume that the operator $A$ generates a quasi-contractive semigroup $S(t)$. Our objective is to guarantee the fixed-time stabilization of the system by means of a suitably designed feedback control strategy. To this end, we construct a nonlinear control law that is specifically devised to suppress the destabilizing effects inherent in the system dynamics and to enforce convergence of the closed-loop trajectory towards the desired equilibrium within a fixed-time.  Moreover, under this feedback law, the closed-loop system is well-posed in the sense that it admits a unique weak solution. More precisely, we propose the following nonlinear feedback:
\begin{equation}\label{15}
     u(y(t))= -C(B^{*}y(t)),
\end{equation}
where $C: \mathcal{U}\rightarrow \mathcal{U}$ is defined by
$$
C(z)= \begin{cases}\left(\nu+\|z\|_{\mathcal{U}}^{-\mu}+\|z\|_{\mathcal{U}}^{\mu} \right)z & \text { if } z \neq 0 \\ 0 & \text { otherwise }\end{cases}
$$
for $0<\mu<1$ and some appropriate $\nu\in \mathbb{R}^{+}$. 
Substituting the   control law (\ref{15}) into system (\ref{eq:system_nonlinear}), we obtain the following closed-loop system:
\begin{eqnarray}\label{systSresule}
	\begin{cases}
		\frac{d }{d t} y(t)-A y(t)+ F(y(t)) + G(y(t))=0, \quad    \hspace*{0.1cm} t>0,\\
		y(0)=y_{0},    \\
	\end{cases}
\end{eqnarray}
where  $G(y)=B C(B^{*}y)$ and $F: \mathcal{H}\rightarrow \mathcal{H} $ is  defined by $F(y)=-f(y)y$.
\begin{remark}
     The proposed control law in \eqref{15} is non-singular because \(\lim\limits_{t\to T} y(t) = 0\) and \(\lim\limits_{y(t) \to 0} \|u(y(t))\| = 0\).  In effect,  
$$\begin{aligned}
 &\lim\limits_{y(t) \to 0} \|u(y(t))\|\\
 =& \lim\limits_{y(t) \to 0} -\left(\left\|B^{*}y(t)\right\|_{\mathcal{U}}^{-\mu}+\left\|B^{*}y(t)\right\|_{\mathcal{U}}^{\mu}+\frac{\omega}{\beta^{2}}\right)\left\|B^{*}y(t)\right\|_{\mathcal{U}}\\
 = &\lim\limits_{y(t) \to 0} -\left(\left\|B^{*}y(t)\right\|_{\mathcal{U}}^{1-\mu}+\left\|B^{*}y(t)\right\|_{\mathcal{U}}^{1+\mu}+\nu \left\|B^{*}y(t)\right\|_{\mathcal{U}}\right).
\end{aligned}
$$  
Using the fact that $\mu \in (0,1)$, then \(\lim\limits_{y(t) \to 0} \|u(y(t))\| = 0\).
\Endofremark
\end{remark}
\subsubsection{Well-posedness}
We begin with the following result which   introduces the existence and uniqueness of the weak  solution of the
closed-loop system  (\ref{systSresule}). 
\begin{theorem}[Well-posedness]\label{thm:M4}  Let $A$  generate a quasi-contractive $C_0$-semigroup $S(t)$  of type  $\omega\geq 0$ on $\mathcal{H}$ and let   the assumptions \Ass{1} hold. Then,  for all $y_{0}\in \mathcal{H}$,  the system (\ref{systSresule}) has a unique weak solution $y\in C([0,+\infty), \mathcal{H})$. Furthermore,   for every $y_0 \in D(A)$, the solution $y$ is  strong  and satisfies:\\
- $y(t) \in D(A)$ for all $t \in[0,+\infty)$,\\
- $y$ is Lipschitz on $[0,+\infty)$,\\
- $\frac{d }{d t}y(t)-A y(t)+ F(y(t)) + G(y(t))=0$ for  a.e $t > 0$. 
\end{theorem}
\subsubsection{Fixed-time stabilization}
We can state our main result on fixed-time stabilization of system (\ref{eq:system_nonlinear}) as follows.
\begin{theorem}[Global fixed-time stability]\label{thm:M5} Assume that the conditions of Theorem  \ref{thm:M4} are satisfied
and suppose  that \Ass{2}  holds. Then,  the system  (\ref{eq:system_nonlinear})  is globally fixed-time stable under the following control:
\begin{equation}\label{systSbnbbbbkamalfenzan}
u(y(t))= \begin{cases}-\left(\left\|B^{*}y(t)\right\|_{\mathcal{U}}^{-\mu}+\left\|B^{*}y(t)\right\|_{\mathcal{U}}^{\mu}+\frac{\omega}{\beta^{2}}+\kappa\right)B^{*}y(t),  
\\  \quad \quad \text { if } y(t) \neq 0, \\ 0, \quad \text { otherwise },\end{cases}
\end{equation}
where  $\mu \in(0,\;1)$. The settling time admits the estimate  $T\leq\frac{\pi}{2\beta^{2} \mu}$.
\end{theorem}
\begin{remark}
The estimation of the settling time is first established for finite-dimensional systems, which is also referred to as predefined-time convergence~\cite{jimenez2020lyapunov}.
\Endofremark
\end{remark}

\section{Proofs}\label{sec:Proofs}
\subsection{Proof of Theorem~\ref{thm:M1}}
\begin{proof}
Note that in this case, due to the absence of the Lipschitz continuity assumption, 
the arguments employed in \cite{polyakov2024finite} or the classical theory developed in \cite{curtain2016stabilization} and \cite{curtain2020introduction}  cannot be directly applied. Instead, our analysis relies on the maximal monotone 
property of the operator and makes use of nonlinear semigroup theory, together with 
the well-posedness framework for abstract evolution equations as established in 
\cite{brezis1973operateurs}.\\
Consider the mapping $g=N \circ B^{*}$, where $N(y)=\|y\|$, for all $y\in \mathcal{H}$. Indeed, it can be easily seen that the subgradient of the operator $N$ is given by $\partial N(y)=\operatorname{sign}(y)$. Furtherome,  since $N$ is a continuous proper and convex function on $\mathcal{H}$  and $B^{*}\in \mathcal{L}(\mathcal{H},\mathcal{U})$, then $g$ is also a continuous proper and convex  function. Moreover,  by Theorem \ref{C1}, its subgradient  $\partial g$  is maximal monotone on $\mathcal{H}$ with $D(\partial g)=\mathcal{H}$. On the other hand, by using (\cite{clarke2013functional}, Theorem 4.13), we obtain $\partial g= B~ \partial N(B^{*}y)= B~\operatorname{sign}(B^{*}y)$.
\\
Now, we define the function $L(y)=\frac{\lambda}{2}\left\|B^{*}y\right\|_{\mathcal{U}}^{2}$.  Then, it can be easily verified that the Fréchet derivative of $L$ is given by $L^{\prime}(y) . h=\lambda\langle BB^{*}(y), h\rangle$, for all $h \in \mathcal{H}$.   Hence, $\lambda~  BB^{*}$ is a maximal monotone operator.
 Furthermore, the operator $A$ generates a quasi-contractive semigroup, then $-A+\omega I$ is maximal monotone  operator. By applying Theorem~\ref{C2} and Theorem~\ref{C3}, we then deduce that the operator  $\Gamma=-A+ \omega~ I + \lambda~  BB^{*}+B~ \operatorname{sign}(B^{*})$ is maximal monotone. As a consequence of  Theorems \ref{C4} and \ref{C5}, the feedback system {\color{black} (\ref{KAMAKkbbb})} admits a unique  weak solution  $y\in C([0,+\infty), \mathcal{H})$.
\end{proof}
\subsection{Proof of Theorem~\ref{thm:M2}}
\begin{proof}
Under assumption \Ass{2}, system (\ref{eq:Abstract}) with control (\ref{con})  is equivalent to system~\eqref{KAMAKkbbb}. Therefore, by Theorem~\ref{thm:M4}, system~\eqref{KAMAKkbbb} admits a unique weak solution for every initial datum \(y_0\in\mathcal H\). Hence, there is  a sequence $y_{0}^n$ in $D(A)$ such that $y_{0}^n \rightarrow y_{0}$, and the system  (\ref{KAMAKkbbb}) with $y_{0}^n$ as initial state possesses a strong solution $y^n \in C([0,+\infty), \mathcal{H}$ verifying $y^n \rightarrow y$ uniformly on $[0, T^1]$ (for  any $T^1>0$ ) as $n \rightarrow \infty$.

 Then,  for a.e $t>0$,  we have $y^n(t) \in D(A)$  and   
\[
\frac{\mathrm{d}}{\mathrm{d} t} y^n(t) \in 
Ay^n(t) - \frac{\omega}{\beta^2} \, BB^{*} y^n(t) 
- \rho \, B \operatorname{sign}\!\bigl(B^{*} y^n(t)\bigr) + \eta(t).
\]
Now, in order to  prove the finite-time stability of the system   (\ref{KAMAKkbbb}). We consider the Lyapunov function candidate  defined by $V(y)=\left\|y\right\|^{2} $, for all $y\in \mathcal{H}$. For a.e. $t >0$, the time derivative of $V$ along the trajectories  of the system  (\ref{KAMAKkbbb}), yields
$$
\begin{aligned}
\frac{1}{2} \frac{d}{d t}\left\|y^n(t)\right\|^{2} &=\langle y^n(t), Ay^n(t)\rangle - \langle \frac{\omega}{\beta^2} BB^{*}y^n(t), y^n(t)  \rangle\\
& - \langle \rho  B(r^n(t)), y^n(t)  \rangle+\langle \eta(t), y^n(t)  \rangle, \\
\end{aligned}
$$
where  $r^n(t)\in \operatorname{sign}(B^{*}y^n(t))$.
By Assumption  \Ass{2} and using the fact that $\langle y^n(t), Ay^n(t)\rangle\leq \omega \|y^n(t)\|^{2}$ , we find
$$
\begin{aligned}
\frac{1}{2} \frac{d}{d t}\|y^n(t)\|^{2}\leq 
\begin{cases}
		 -\rho \beta \|y^n(t)\|+ \|y^n(t)\| \|\eta(t)\|, &  y^n(t) \neq 0, \\
		0, &  y^n(t)=0.
	\end{cases}
\\
\end{aligned}
$$
Under the Assumption \Ass{3}, we have 
$$
\begin{aligned}
\frac{1}{2} \frac{d}{d t}\|y^n(t)\|^{2}\leq 
\begin{cases}
		-( \rho \beta -M) \|y^n(t)\|, &  y^n(t) \neq 0, \\
		0, &  y^n(t)=0.
	\end{cases}
\\
\end{aligned}
$$
Hence,  according to the comparison principle, we have
$$ 
	\begin{cases}
	\begin{aligned}
		& \|y^n(t)\|\leq \|y_{0}^n\| -  ( \rho \beta -M)   t, & \text{ if }   t \leq \frac{\|y_{0}^n\|}{ \rho \beta -M},\\
			& y^n(t)=0, & \text{ if }  t \geq  \frac{\|y_{0}^n\|}{ \rho \beta -M}.
		\end{aligned}
	\end{cases} 
$$
Then, letting  $n\rightarrow +\infty$,  we deduce that
$$ 
	\begin{cases}
	\begin{aligned}
		& \left\|y(t)\right\|\leq \left\|y_{0}\right\| -  ( \rho \beta -M) t,   & \text{ if }   t \leq \frac{\left\|y_{0}\right\|}{  \rho \beta -M},\\
			& y(t)=0,  &\text{ if }  t \geq  \frac{\left\|y_{0}\right\|}{  \rho \beta -M}.
		\end{aligned}
	\end{cases} 
$$
This achieves the proof.\\
\end{proof}
\subsection{Proof of Theorem~\ref{thm:M3}}
\begin{proof}
    \textbf{Step 1: Well-posedness result:}\\
We define, for every $y \in \mathcal{H}$, 
\[
L(y)=\|B^{*}y\|_{\mathcal{U}}^{\zeta}\,BB^{*}y.
\]
It is easy to verify that $L$ coincides with the Fréchet derivative of the following convex continuous functional:
\[
g(y)=\frac{\big\langle BB^{*}y,\,y\big\rangle^{\,1+\frac{\zeta}{2}}}{2+\zeta}.
\]
Then the operator $\Gamma + L $ is maximal monotone. The proof concludes by following the same steps as in Theorem \ref{thm:M1}.\\
\textbf{Step 2: Fixed time stabilization:}\\
By the same  approximation argument used in Theorem \ref{thm:M2}, we may,  without loss of generality, assume that  the unique  solution $y$ of (\ref{eq:Abstract})-(\ref{con2}) is strong.   Then, for a.e. $t > 0$,  we have $y(t)\in D(A)$ and  the solution satisfies 
$$ 
\begin{aligned}
\frac{\mathrm{d}}{\mathrm{d} t}y(t)\in Ay(t)   &-\left(\frac{\omega}{\beta^2} +\left\|B^{*}y(t)\right\|_{\mathcal{U}}^{\zeta}\right)BB^{*}y(t)\\&-\rho~ B\operatorname{sign}(B^{*}y(t))+ \eta(t),
\end{aligned}
$$
Now, in order to  prove the finite-time stability of the system   (\ref{KAMAKkbbb}). We consider the Lyapunov function candidate  defined by $V(y)=\left\|y\right\|^{2} $, for all $y\in \mathcal{H}$. For a.e. $t >0$, the time derivative of $V$ along the trajectories  of the system  (\ref{KAMAKkbbb}), yields
$$
\begin{aligned}
&\frac{1}{2} \frac{d}{d t}\left\|y(t)\right\|^{2} = -\left(\frac{\omega}{\beta^2} +\left\|B^{*}y(t)\right\|_{\mathcal{U}}^{\zeta}\right)\langle  BB^{*}y(t), y(t)  \rangle\\
& - \langle \rho  B(r(t)), y(t)  \rangle  +\langle \eta(t), y(t)  \rangle+\langle y(t), Ay(t)\rangle,  \\
\end{aligned}
$$
where $r(t)\in \operatorname{sign}(B^{*}y(t))$. Since the operator $A-\omega I$  is dissipative, and under assumptions \Ass{2} and  \Ass{3},   it follows that
$$
\begin{aligned}
\frac{1}{2} \frac{d}{d t}\|y(t)\|^{2}\leq 
\begin{cases}
		-( \rho \beta -M) \|y(t)\|-\beta^{\zeta+ 2} \|y(t)\|^{\zeta+ 2}  \\  \quad\quad\quad \text { if} ~ ~   y(t) \neq 0, \\
		0,    \quad\quad \text { if} ~ ~ y(t)=0.
	\end{cases}
\\
\end{aligned}
$$
Following the approach  in Lemma $1$ of \cite{polyakov2011nonlinear},  the closed-loop system is fixed-time stable and the settling time satisfies (\ref{T}).
\end{proof}

\subsection{Proof of Theorem~\ref{thm:M4}}
\begin{proof}
To establish the well-posedness of  (\ref{systSresule}), we'll show that the operator $\mathcal{B}=-A+F_{1}+F_{2}+G+(\omega+ \kappa )I$ is maximal monotone, where $\kappa$ is the Lipschitz constant of $F_{2}$. We first show that the operator 
\[
-A+F_{1}+F_{2}+(\omega+ \kappa )I : D(A) \rightarrow \mathcal{H}
\]
is maximal monotone. For any $y_1, y_2 \in \mathcal{H}$, we compute
$$
\begin{aligned}
& \left\langle F_2(y_1) + \kappa y_1 - F_2(y_2) - \kappa y_2,\, y_1 - y_2 \right\rangle \\
& \quad = \left\langle F_2(y_1) - F_2(y_2),\, y_1 - y_2 \right\rangle 
      + \kappa \|y_1 - y_2\|^2 \\
& \quad \geq - \|F_2(y_1) - F_2(y_2)\| \, \|y_1 - y_2\|
      + \kappa \|y_1 - y_2\|^2.
\end{aligned}
$$
Using Assumption~\Ass{1}, we deduce that
\[
\left\langle F_2(y_1) + \kappa y_1 - F_2(y_2) - \kappa y_2,\, y_1 - y_2 \right\rangle \geq 0,
\]
so that $F_2 + \kappa I$ is a maximal monotone operator.  
Since $A$ generates a quasi-contractive semigroup, the operator $-A + \omega I$ is  maximal monotone.  
Therefore, combining this with the maximal monotonicity of $F_2$ (Assumption \Ass{1}), we conclude, by Theorems \ref{C2} and \ref{C3},  that the sum 
\[
-A + F_1 + F_2 + (\omega + \kappa) I
\]
is maximal monotone as well.\\
To complete the proof, it remains to show that the operator $G$ is maximal monotone. 
More precisely, we will prove that $G$ can be expressed as the Fréchet derivative of a 
lower semi-continuous convex functional $\Phi$, that is, $\Phi' = G$. \\ 
Let us therefore introduce the functional $\Phi : \mathcal{H} \longrightarrow \mathbb{R}^{+}$ defined by
$$\begin{aligned}
\Phi(y)&=\frac{\left\|B^{*}y\right\|_{\mathcal{U}}^{2-\mu}}{2-\mu}+\frac{\left\|B^{*}y\right\|_{\mathcal{U}}^{2+\mu}}{2+\mu}+\frac{\nu}{2}\left\|B^{*}y\right\|_{\mathcal{U}}^{2}\\
&=\frac{\nu}{2 }\langle BB^{*} y, y\rangle+\frac{\langle BB^{*} y, y\rangle^{1-\frac{\mu}{2}}}{2-\mu}+\frac{\langle BB^{*} y, y\rangle^{1+\frac{\mu}{2}}}{2+\mu}.
\end{aligned}
$$
Clearly, $\Phi$ is convex and continuous on $\mathcal{H}$. 
 Now, writing  $\Phi=\Lambda\circ\varphi$ with 
 $$
 \Lambda(s)=\frac{\nu}{2 }s+\frac{s^{2-\mu}}{2-\mu}+\frac{s^{2+\mu}}{2+\mu}, \forall s\in \mathbb{R}^{+}.
 $$
Let $\varphi(y)=\langle BB^{*}y,\,y\big\rangle$, for all $y\in \mathcal{H}$. It follows that the Fréchet derivative of $\Phi$ at $y$ such that $B^{*}y\neq0$ is given by
$$
\begin{aligned}
\Phi^{\prime}(y) &= \frac{\nu}{2 }\varphi^{\prime}(y)+\Lambda^{\prime}(\varphi(y)) \varphi^{\prime}(y)\\ 
&=\left(\nu+\left\|B^{*}y\right\|_{\mathcal{U}}^{-\mu}+\left\|B^{*}y\right\|_{\mathcal{U}}^{\mu}\right)BB^{*}y\\&=BC(B^{*}y)
\\&= G(y).
\end{aligned}
$$
In addition, for $B^{*}y=0$, using   $0<\mu<1$, we have
\[
\begin{aligned}
\lim _{\|h\| \rightarrow 0} \frac{\Phi(y+h)}{\|h\|} 
&= \lim _{\|h\| \rightarrow 0} \Bigg(
    \frac{\nu}{2\|h\|}\langle BB^{*} h, h\rangle \\
&\quad + \frac{\langle BB^{*} h, h\rangle^{1-\frac{\mu}{2}}}{(1-\mu)\|h\|} 
    + \frac{\langle BB^{*} h, h\rangle^{1+\frac{\mu}{2}}}{(1+\mu)\|h\|} 
\Bigg) \\
&= 0
\end{aligned}
\]
In other words, for all $y \in \operatorname{Ker} B^{*} $ the Fréchet derivative is given by $\Phi^{\prime}(y)=0$, and therefore $\Phi^{\prime}=G$. Then, by Theorem~\ref{C1}, the operator $G$ is maximal monotone. Consequently, applying Theorem~\ref{C3}, we deduce that $\mathcal{B}$, being the sum of maximal monotone operators, is itself a maximal monotone operator.\\
  Then, according to Theorems   \ref{C4} and \ref{C5},
  we  conclude that for all $y_{0} \in \mathcal{H}$,   the system  (\ref{systSresule}) has a unique weak solution $y\in C([0,+\infty), \mathcal{H})$.
\end{proof}

\subsection{Proof of Theorem~\ref{thm:M5}}
\begin{proof}
Observing that system (\ref{eq:system_nonlinear}), when endowed with the feedback control law (\ref{systSbnbbbbkamalfenzan}), is equivalent to system  (\ref{systSresule}) with the parameter identification $\nu = \kappa+\tfrac{\omega}{\beta^{2}}$, we can exploit the analytical framework  established in Theorem \ref{thm:M4}. In particular, according to  this theorem,  the system   (\ref{systSresule}) admits a unique weak solution  for all $y_{0}\in \mathcal{H}$. Hence, there is  a sequence $y_{0}^n$ in $D(A)$ such that $y_{0}^n \rightarrow y_{0}$, and the system  (\ref{systSresule}) with $y_{0}^n$ as initial state possesses a strong solution $y^n \in C([0,+\infty), \mathcal{H}$ verifying $y^n \rightarrow y$ uniformly on $[0, T^1]$ (for  any $T^1>0$ ) as $n \rightarrow \infty$.
 Then,  for a.e $t>0$,  we have $y^n(t) \in D(A)$  and   
$$ 
\begin{aligned}
\frac{\mathrm{d}}{\mathrm{d} t}y^n(t)-Ay^n(t)+ F(y^n(t)) +G(y^n(t)) =0,
\end{aligned}
$$
Now, in order to  prove the fixed-time stability of the system   (\ref{systSresule}). We consider the Lyapunov function candidate  defined by $V(y)=\left\|y\right\|^{2} $, for all $y\in \mathcal{H}$. Our objective is to analyze the time derivative of $V$ along the trajectories of the closed-loop system  and to verify that it satisfies an inequality of the form (\ref{fix-time}). \\
For a.e. $t >0$, the time derivative of $V$ along the trajectories  of the system  (\ref{systSresule}), yields
$$
\begin{aligned}
\frac{1}{2} \frac{d}{d t}\left\|y^n(t)\right\|^{2} &=\langle y^n(t), Ay^n(t)-F(y^n(t))-G(y^n(t))\rangle. \\
\end{aligned}
$$
Using Assumption \Ass{2}, we can see that 
\begin{equation}\label{iné}
    \langle f(y)y, y\rangle= - \langle F(y), y\rangle \leq \kappa \|y\|^2, \quad \text{for all } y\in \mathcal{H}.
\end{equation}
By Assumptions \Ass{1}, \Ass{2},  and  (\ref{iné}), we find
\begin{gather*}
\frac{1}{2} \frac{d}{dt} V(y^n(t)) \leq
-\beta^{2-\mu} V(y^n(t))^{1-\frac{\mu}{2}} - \beta^{2+\mu} V(y^n(t))^{1+\frac{\mu}{2}},\\
\quad y^n(t) \neq 0,  \quad \text{and}\\
 \frac{1}{2} \frac{d}{dt} V(y^n(t)) \leq 0, \quad y^n(t) = 0.
\end{gather*}
Hence,  by following the approach of
Lemma $2$ of \cite{parsegov2012nonlinear}, we obtain 
	\begin{gather*}
			 \arctan (\beta^{\mu} V(y^n(t))^{\frac{\mu}{2}})\leq \arctan (\beta^{\mu}V(y_{0}^n)^{\frac{\mu}{2}}) -  \beta^{2}\mu  t,  \\\text{ if }   t \leq T_1, \quad \text{and}\\
			 \arctan (\beta^{\mu} V(y^n(t))^{\frac{\mu}{2}})=0, \text{ if }  t \geq  T_1,
			\end{gather*}

where $T_1=\frac{\arctan (\beta^{\mu}V(y_{0}^n)^{\frac{\mu}{2}}) }{ \beta^{2}\mu}$. Then, letting  $n\rightarrow +\infty$,  we deduce that
$$ 
	\begin{cases}
	\begin{aligned}
		& \arctan(\beta^{\mu} \left \|y(t)\right\|^{\mu})\leq \arctan(\beta^{\mu}\left\|y_{0}\right\|^{\mu}) -  \beta^{2}\mu  t,  \text{ if }   t \leq T,\\
		& \arctan(\beta^{\mu}\left \|y(t)\right\|^{\mu})=0,  \text{ if }  t \geq T,
		\end{aligned}
	\end{cases} 
$$
where $T=\frac{\arctan(\beta^{\mu}\left\|y_{0}\right\|^{\mu})}{ \beta^{2}\mu}$.
Hence, we deduce that $y(t)=0$, for all $t\geq T$ and $T\leq \frac{\pi}{2\mu \beta^{2}}.$ This achieves the proof.
\end{proof}

\section{Application: Heat equation }\label{sec:applications}
Let $\Omega=(0,1)$  and let us consider the following   mono-dimensional heat equation:
\begin{IEEEeqnarray}{c}\label{exemple1}
\left\{
\begin{aligned}
&y_t(x, t) = \Delta y(x, t) + R(y(x, t),t) \\
& + \sqrt{a(x)} \, u(y(x, t)), \quad (x, t) \in \Omega \times (0,+\infty) \\
&y(x, t) = 0, \quad (x, t) \in \partial \Omega \times (0,+\infty) \\
&y(x, 0) = y_0, \quad x \in \Omega
\end{aligned}
\right.
\end{IEEEeqnarray}
where  $y (t)=y (\cdot,t) $ is the state, $a \in L^{\infty}(\Omega)$ and $a(x) \geq c$ a.e. $x \in \Omega$ for some $c>0$.\\
We can write  (\ref{exemple1}) in the form of  (\ref{eq:system_nonlinear}) on $L^{2}(\Omega )$  if we set $Ay=\Delta y$ for $y\in D(A)=\mathcal{H}_{0}^{1}(\Omega ) \cap \mathcal{H}^2(\Omega )$,  the spectrum of $A$ is given by the
simple eigenvalues $\lambda_{j}= -(j\pi)^{2}$ and eigenfunctions $\phi_{j}=\sqrt 2~  \sin(j\pi x)$, $j\geq 1$. Furthermore, $A$ generates a contraction semigroup  given by $S(t) y=\sum\limits_{j=1}^{\infty} \exp \left(\lambda_{j} t\right)\langle y, \phi_j\rangle \phi_j$, for all $y\in L^{2}(\Omega )$ and all $t\geq 0$. In particular, the semigroup $(S(t))_{t \geq 0}$ is of type $\omega = 0$. Moreover, \(R\) denotes the nonlinearity operator, or equivalently a matched disturbance, acting on \(\mathcal{H}\).\\
Consider the bounded operator $B :L^{2}(\Omega) \rightarrow L^{2}(\Omega)$ defined by $Bz= \sqrt{a(.)}\hspace*{0.1cm}z$. Then, the assumption \Ass{2} is verified with $\beta=\sqrt{c}$.  \\
The problem of distributed finite-time control for the  heat system (\ref{exemple1})  with $f=0$ has been extensively studied in  \cite{sob}, while the specific case where $a(x)=1$ has been considered in  \cite{ouzahra2021finite}. 
In this work, we further extend the analysis by addressing the problem of finite- and fixed-time stabilization of system~\eqref{exemple1} in the presence of disturbances.
\subsection{ Case 1: $R(y,t)=\sin(t) \cos(x)$, for all $y\in \mathcal{H} $ and $t\geq 0$}
In this case, the system \eqref{exemple1} can be expressed in the abstract form \eqref{eq:Abstract} with
\[
\eta(t) = \sin(t) \cos(x).
\]
Then, by virtue of  Theorem \ref{thm:M1}, it follows that the system \eqref{exemple1} is finite-time stable under the following control:
\begin{gather}\label{exemple111}
u(y (\cdot,t)) = \begin{cases} 
-\frac{\rho\sqrt{a(\cdot)} y (\cdot,t)}{\|\sqrt{a(\cdot)} y (\cdot,t)\|}
& \text{if } y (\cdot,t) \neq 0, \\ 
\rho\mathcal{B}_{\mathcal{H}} & \text{otherwise.}
\end{cases}
\end{gather}
where  $\mu \in(0,\;1)$.  Moreover, according to  Theorem \ref{thm:M3}, we deduce that the system \eqref{exemple1} is fixed-time stable under the following control:
\begin{gather}\label{exemple1111}
u(y (\cdot,t)) = \begin{cases} 
\begin{aligned}[t]
&-\frac{\rho\sqrt{a(\cdot)} y (\cdot,t)}{\|\sqrt{a(\cdot)} y (\cdot,t)\|}\\
& - \left\|\sqrt{a(\cdot)} y (\cdot,t)\right\|^{\zeta}\sqrt{a(\cdot)} y (\cdot,t), \quad  \text{if } y (\cdot,t) \neq 0,
\end{aligned}   \\ 
\rho\mathcal{B}_{\mathcal{H}}, \quad  \text{if } y (\cdot,t) = 0,
\end{cases}
\end{gather}
where $\zeta > 0 $.
\subsection{ Case 2: $R(y,t)=f(y(x, t))y(x, t)$ for some appropriate function $f$}
In this case, we consider the function
\[
f(y)= -\frac{\|\sqrt{a(\cdot)}\,y\|^{- \mu}\,a(\cdot)}{1+\|\sqrt{a(\cdot)}\,y \|^{2}} .
\]
The finite-time stabilization of system~(\ref{exemple1}) with this choice of \(f\) was established in \cite{fenza2025finite}.
In the present work, we address the fixed-time stabilization of the same system.
To this end, we apply Theorem \ref{thm:M5}. Moreover, Assumption~\Ass{1} is automatically satisfied.
Consequently, system~(\ref{exemple1}) is fixed-time stable under the following control law:
\begin{gather}\label{exemple112}
u(y (\cdot,t)) = 
-\left(\|\sqrt{a(\cdot)} y (\cdot,t)\|^{- \mu}+\|\sqrt{a(\cdot)} y (\cdot,t)\|^{ \mu}\right) \nonumber \\
\quad \times \sqrt{a(\cdot)} y (\cdot,t)
 \quad \text{if } \quad y (\cdot,t) \neq 0, \quad \text{and} \\ 
u(y (\cdot,t)) =  0 \quad \text{if } \quad y (\cdot,t) = 0,\nonumber
\end{gather}
where  $\mu \in(0,\;1)$.

\subsection{Simulations}
In this section, we present numerical simulations that illustrate the theoretical results established in the previous sections. 
Figure~1 shows the evolution of the state of system~(\ref{exemple1}) under the nonlinear control law \(u(y(\cdot,t))\) defined in (23), 
for the case \(a(x)=x^{2}+0.01\), \(\mu=\tfrac{1}{2}\), and the initial condition \(y_0(x)=\sin(\pi x)\), for all \(x\in \Omega=(0,1)\). 
Figures~2 and~3 display, respectively, the distributed control law \(u(y(\cdot,t))\) and its norm \(\|u(y(\cdot,t))\|\) in the fixed-time stabilization case. 
Finally, Figure~4 illustrates the evolution of the state norm \(\|y(t)\|\) for different choices of the initial condition \(y_0\). 
These results clearly show that the settling time is uniformly upper bounded by a constant independent of the initial conditions, 
which confirms that the convergence time does not depend on the choice of \(y_0\).
The considered initial states are given by
\[
y_0^1 (x)= \sin(\pi x), \quad 
y_0^2(x) = \sin(2\pi x), \quad 
\]
\[
y_0^3(x) = x(1-x), \quad 
y_0^4 (x)= \exp\!\bigl(-5(x-0.5)^{2}\bigr).
\]
\begin{figure}[h!]
	\centerline{\includegraphics[scale=0.7,width=90mm]{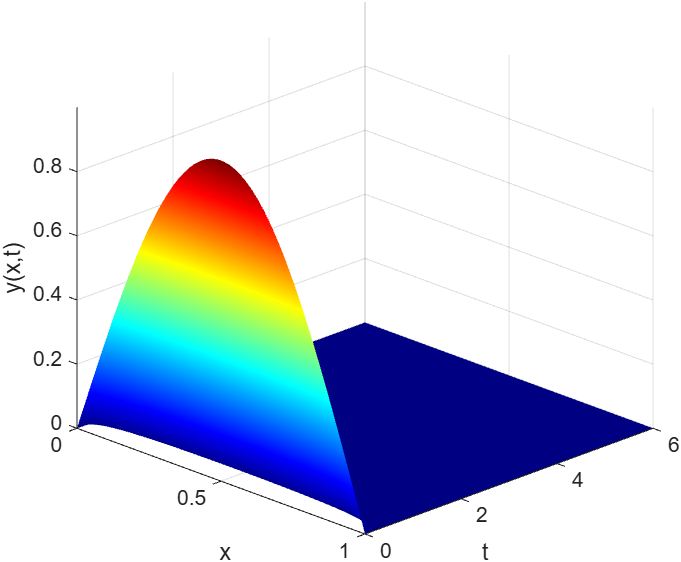}}
    	\caption{ The evolution of the state  $y(\cdot,t)$  of the  system (\ref{exemple1})   with the control $u(y(\cdot,t))$  in (23).  } 
	\end{figure}
\begin{figure}[h!]
	\centerline{\includegraphics[scale=0.7,width=90mm]{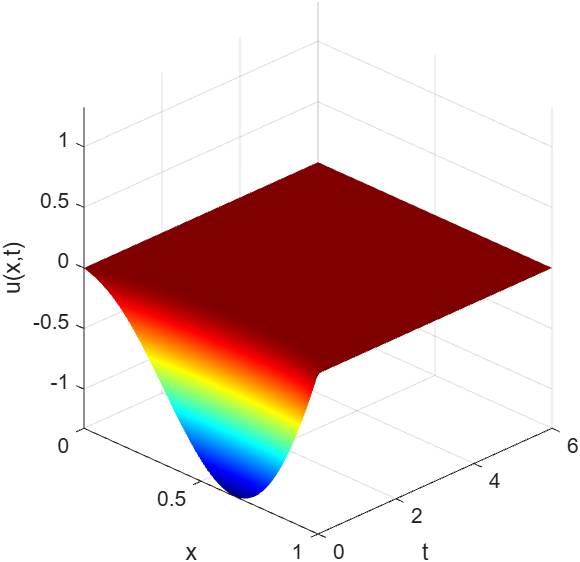}}
    	\caption{  The evolution of the nonlinear control $u(y(\cdot,t))$ given in (23).     } 
	\end{figure}

\begin{figure}[h!]
	\centerline{\includegraphics[scale=0.7,width=90mm]{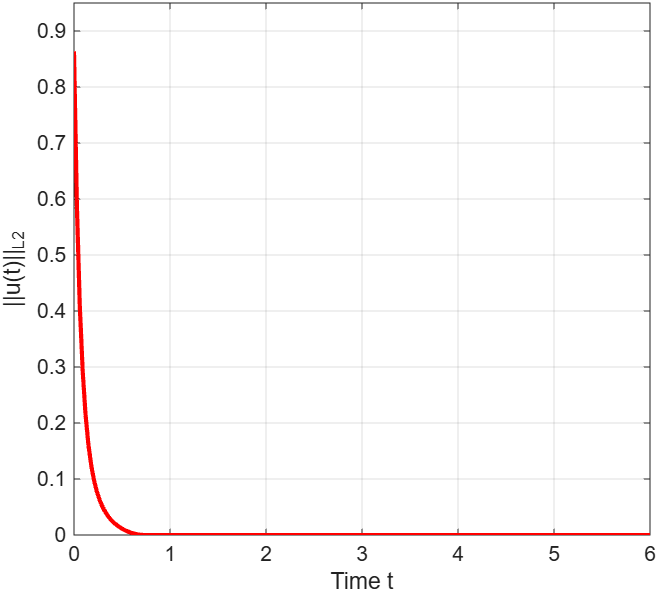}}
    	\caption{Evolution of $\|u(y(\cdot,t))\|$   } 
	\end{figure}
	\begin{figure}[h!]
		\centerline{\includegraphics[scale=0.7,width=90mm]{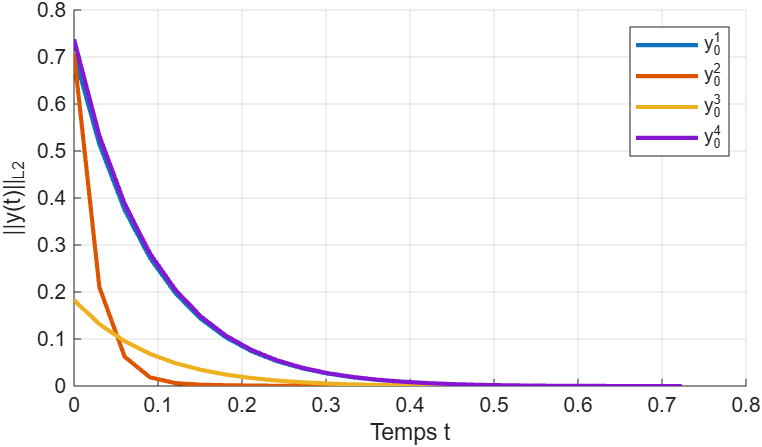}
        }
        \caption{The evolution of $\|y(\cdot,t)\|$ for different values  of $y_0$ }
	\end{figure}

\section{Conclusions and Future Work}\label{sec:conclusions}

In this paper, we addressed the stabilization of two classes of abstract systems subject to perturbations and nonlinearities. Both finite- and fixed-time stability properties were established. For the first class (linear abstract systems), we considered persistent perturbations, which required a discontinuous control law; the well-posedness of the closed-loop system was rigorously proved using maximal monotone operator theory. For the second class (nonlinear abstract systems), we established fixed-time stability and well-posedness of the solutions, again using maximal monotone operator theory. The theoretical results were illustrated through applications to heat equations, including numerical simulations.

For future work, the assumptions on the nonlinear functions can be further relaxed, and alternative approaches, such as fixed-point theory, could be explored to establish well-posedness.

\bibliographystyle{IEEEtran}
  
\bibliography{Bib}  

\begin{thebibliography}{10}
\providecommand{\url}[1]{#1}
\csname url@samestyle\endcsname
\providecommand{\newblock}{\relax}
\providecommand{\bibinfo}[2]{#2}
\providecommand{\BIBentrySTDinterwordspacing}{\spaceskip=0pt\relax}
\providecommand{\BIBentryALTinterwordstretchfactor}{4}
\providecommand{\BIBentryALTinterwordspacing}{\spaceskip=\fontdimen2\font plus
\BIBentryALTinterwordstretchfactor\fontdimen3\font minus
  \fontdimen4\font\relax}
\providecommand{\BIBforeignlanguage}[2]{{%
\expandafter\ifx\csname l@#1\endcsname\relax
\typeout{** WARNING: IEEEtran.bst: No hyphenation pattern has been}%
\typeout{** loaded for the language `#1'. Using the pattern for}%
\typeout{** the default language instead.}%
\else
\language=\csname l@#1\endcsname
\fi
#2}}
\providecommand{\BIBdecl}{\relax}
\BIBdecl

\bibitem{ma2025integral}
L.~Ma, V.~Andrieu, D.~Astolfi, M.~Bajodek, C.-Z. Xu, and X.~Lou, ``Integral
  action feedback design for conservative abstract systems in the presence of
  input nonlinearities,'' \emph{IEEE Transactions on Automatic Control}, 2025.

\bibitem{Fuller1960}
A.~Fuller, ``Relay control systems optimized for various performance
  criteria,'' in \emph{Proceedings of the 1st IFAC Triennial World Congress},
  1960, pp. 510--519.

\bibitem{baji2007asymptotics}
B.~Baji, A.~Cabot, and J.~D{\'\i}az, ``Asymptotics for some nonlinear damped
  wave equation: finite time convergence versus exponential decay results,'' in
  \emph{Annales de l'Institut Henri Poincar{\'e} C, Analyse non lin{\'e}aire},
  vol.~24, no.~6.\hskip 1em plus 0.5em minus 0.4em\relax Elsevier, 2007, pp.
  1009--1028.

\bibitem{Brogliato1999}
B.~Brogliato, \emph{Nonsmooth Mechanics}, 2nd~ed., ser. Communications and
  Control Engineering Series.\hskip 1em plus 0.5em minus 0.4em\relax London:
  Springer, 1999.

\bibitem{pohjolainen1982robust}
S.~Pohjolainen, ``Robust multivariable pi-controller for infinite dimensional
  systems,'' \emph{IEEE Transactions on Automatic Control}, vol.~27, no.~1, pp.
  17--30, 1982.

\bibitem{Haraux1978Thesis}
A.~Haraux, ``Op{\'e}rateurs maximaux monotones et oscillations forc{\'e}es non
  lin{\'e}aires,'' Ph.D. dissertation, Universit{\'e} Pierre et Marie Curie,
  Paris, 1978.

\bibitem{Cabannes1978}
H.~Cabannes, ``Mouvement d’une corde vibrante soumise {\`a} un frottement
  solide,'' \emph{C. R. Acad. Sci. Paris S{\'e}r. A-B}, vol. 287, pp. 671--673,
  1978.

\bibitem{Cabannes1981}
------, ``Study of motions of a vibrating string subject to solid friction,''
  \emph{Math. Methods Appl. Sci.}, vol.~3, pp. 287--300, 1981.

\bibitem{CoronNguyen2022}
J.-M. Coron and H.-M. Nguyen, ``{Lyapunov} functions and finite-time
  stabilization in optimal time for homogeneous linear and quasilinear
  hyperbolic systems,'' \emph{Annales de l'Institut Henri Poincaré C, Analyse
  non linéaire}, vol.~39, no.~5, pp. 1235--1260, 2022.

\bibitem{CoronNguyen2020}
------, ``Finite-time stabilization in optimal time of homogeneous quasilinear
  hyperbolic systems in one dimensional space,'' \emph{ESAIM: Control,
  Optimisation and Calculus of Variations}, vol.~26, p.~24, 2020, id/No 119.

\bibitem{Utkin1992}
V.~I. Utkin, \emph{Sliding Modes in Control and Optimization}.\hskip 1em plus
  0.5em minus 0.4em\relax Berlin: Springer, 1992.

\bibitem{orlov2002discontinuous}
Y.~Orlov, ``Discontinuous unit feedback control of uncertain
  infinite-dimensional systems,'' \emph{IEEE transactions on automatic
  control}, vol.~45, no.~5, pp. 834--843, 2002.

\bibitem{cortes2006finite}
J.~Cort{\'e}s, ``Finite-time convergent gradient flows with applications to
  network consensus,'' \emph{Automatica}, vol.~42, no.~11, pp. 1993--2000,
  2006.

\bibitem{coron2017null}
J.-M. Coron and H.-M. Nguyen, ``Null controllability and finite time
  stabilization for the heat equations with variable coefficients in space in
  one dimension via backstepping approach,'' \emph{Archive for Rational
  Mechanics and Analysis}, vol. 225, pp. 993--1023, 2017.

\bibitem{efimov2021finite}
D.~Efimov, A.~Polyakov \emph{et~al.}, ``Finite-time stability tools for control
  and estimation,'' \emph{Foundations and Trends{\textregistered} in Systems
  and Control}, vol.~9, no. 2-3, pp. 171--364, 2021.

\bibitem{coron2021boundary}
J.-M. Coron, L.~Hu, G.~Olive, and P.~Shang, ``Boundary stabilization in finite
  time of one-dimensional linear hyperbolic balance laws with coefficients
  depending on time and space,'' \emph{Journal of Differential Equations}, vol.
  271, pp. 1109--1170, 2021.

\bibitem{nguyen2024rapid}
H.-M. Nguyen, ``Rapid stabilization and finite time stabilization of the
  bilinear schr\"odinger equation,'' \emph{arXiv preprint arXiv:2405.10002},
  2024.

\bibitem{polyakov2016finite}
A.~Polyakov, J.-M. Coron, and L.~Rosier, ``On finite-time stabilization of
  evolution equations: A homogeneous approach,'' in \emph{2016 IEEE 55th
  Conference on Decision and Control (CDC)}.\hskip 1em plus 0.5em minus
  0.4em\relax IEEE, 2016, pp. 3143--3148.

\bibitem{han2024input}
X.-X. Han, D.~Efimov, A.~Polyakov, and K.-N. Wu, ``Input-to-state stability
  analysis of heat equation with boundary finite-time control,''
  \emph{Automatica}, vol. 160, p. 111443, 2024.

\bibitem{10}
A.~Polyakov, D.~Efimov, and W.~Perruquetti, ``Finite-time and fixed-time
  stabilization: Implicit {Lyapunov} function approach,'' \emph{Automatica},
  vol.~51, pp. 332--340, 2015.

\bibitem{15}
A.~Polyakov, J.~M. Coron, and L.~Rosier, ``On finite-time stabilization of
  evolution equations: a homogeneous approach,'' in \emph{Conference on
  Decision and Control}, 2016, pp. 3143--3148.

\bibitem{17}
------, ``On homogeneous finite time control for linear evolution equation in
  hilbert space,'' \emph{IEEE Transactions on Automatic Control}, vol.~63, pp.
  3143--3150, 2018.

\bibitem{ouzahra2021finite}
M.~Ouzahra, ``Finite-time control for the bilinear heat equation,''
  \emph{European Journal of Control}, vol.~57, pp. 284--293, 2021.

\bibitem{fenza2025decomposition}
K.~Fenza, M.~Labbadi, and M.~Ouzahra, ``A decomposition method for finite-time
  stabilization of bilinear systems with applications to parabolic and
  hyperbolic equations,'' \emph{arXiv e-prints}, pp. arXiv--2506, 2025.

\bibitem{sogore2020}
M.~Sogoré and C.~Jammazi, ``On the global finite-time stabilization of
  bilinear systems by homogeneous feedback laws. applications to some
  {PDE}'s,'' \emph{Journal of Mathematical Analysis and Applications}, vol.
  486, no.~2, pp. 123--815, 2020.

\bibitem{sob}
H.~Najib and M.~Ouzahra, ``Output finite-time stabilisation of a class of
  linear and bilinear control systems,'' \emph{International journal of
  control}, vol.~9, pp. 325--334, 2023.

\bibitem{balogoun2023}
I.~Balogoun, S.~Marx, T.~Liard, and F.~Plestan, ``Super-twisting sliding mode
  control for the stabilization of a linear hyperbolic system,'' \emph{IEEE
  Control Systems Letters}, vol.~7, pp. 1--6, 2023.

\bibitem{pilloni2024sliding}
A.~Pilloni, A.~Pisano, E.~Usai, and Y.~Orlov, ``Sliding mode boundary control
  for heat equations with uncertain dynamic actuators,'' in \emph{2024 17th
  International Workshop on Variable Structure Systems (VSS)}.\hskip 1em plus
  0.5em minus 0.4em\relax IEEE, 2024, pp. 93--98.

\bibitem{pisano2012boundary}
A.~Pisano and Y.~Orlov, ``Boundary second-order sliding-mode control of an
  uncertain heat process with unbounded matched perturbation,''
  \emph{Automatica}, vol.~48, no.~8, pp. 1768--1775, 2012.

\bibitem{orlov2004robust}
Y.~Orlov, Y.~Lou, and P.~Christofides, ``Robust stabilization of
  infinite-dimensional systems using sliding-mode output feedback control,''
  \emph{International Journal of Control}, vol.~77, no.~12, pp. 1115--1136,
  2004.

\bibitem{pisano2011tracking}
A.~Pisano, Y.~Orlov, and E.~Usai, ``Tracking control of the uncertain heat and
  wave equation via power-fractional and sliding-mode techniques,'' \emph{SIAM
  Journal on Control and Optimization}, vol.~49, no.~2, pp. 363--382, 2011.

\bibitem{orlov2013boundary}
Y.~Orlov, A.~Pisano, and E.~Usai, ``Boundary control and observer design for an
  uncertain wave process by second-order sliding-mode technique,'' in
  \emph{52nd IEEE Conference on Decision and Control}.\hskip 1em plus 0.5em
  minus 0.4em\relax IEEE, 2013, pp. 472--477.

\bibitem{orlov1995sliding}
Y.~Orlov, ``Sliding mode control in {Banach} space,'' \emph{IFAC Proceedings
  Volumes}, vol.~28, no.~14, pp. 473--476, 1995.

\bibitem{chitour2024sliding}
Y.~Chitour, A.~Dahmani, M.~Labbadi, and C.~Roman, ``Sliding mode observation
  for a 1d wave equation with dynamic boundary conditions.'' in \emph{2024 IEEE
  63rd Conference on Decision and Control (CDC)}.\hskip 1em plus 0.5em minus
  0.4em\relax IEEE, 2024, pp. 1980--1986.

\bibitem{Levaggi2002a}
L.~Levaggi, ``Infinite dimensional systems sliding motions,'' \emph{Eur. J.
  Control}, vol.~8, pp. 508--518, 2002.

\bibitem{Levaggi2002b}
------, ``Sliding modes in {Banach} spaces,'' \emph{Differential and Integral
  Equations}, vol.~15, pp. 167--189, 2002.

\bibitem{bensoussan2007}
A.~Bensoussan, G.~Da~Prato, M.~C. Delfour, and S.~K. Mitter,
  \emph{Representation and Control of Infinite Dimensional Systems}.\hskip 1em
  plus 0.5em minus 0.4em\relax Springer, 2007, vol.~2.

\bibitem{curtain2020}
R.~Curtain and H.~Zwart, \emph{Introduction to Infinite-Dimensional Systems
  Theory: A State-Space Approach}.\hskip 1em plus 0.5em minus 0.4em\relax
  Springer Nature, 2020, vol.~71.

\bibitem{coron2021}
J.-M. Coron, L.~Hu, G.~Olive, and P.~Shang, ``Boundary stabilization in finite
  time of one-dimensional linear hyperbolic balance laws with coefficients
  depending on time and space,'' \emph{Journal of Differential Equations}, vol.
  271, pp. 1109--1170, 2021.

\bibitem{guo2012}
B.-Z. Guo and F.-F. Jin, ``Sliding mode and active disturbance rejection
  control to stabilization of one-dimensional anti-stable wave equations
  subject to disturbance in boundary input,'' \emph{IEEE Transactions on
  Automatic Control}, vol.~58, no.~5, pp. 1269--1274, 2012.

\bibitem{guo2013}
------, ``The active disturbance rejection and sliding mode control approach to
  the stabilization of the euler–bernoulli beam equation with boundary input
  disturbance,'' \emph{Automatica}, vol.~49, no.~9, pp. 2911--2918, 2013.

\bibitem{moreno2012}
J.~A. Moreno and M.~Osorio, ``Strict {Lyapunov} functions for the
  super-twisting algorithm,'' \emph{IEEE Transactions on Automatic Control},
  vol.~57, no.~4, pp. 1035--1040, 2012.

\bibitem{pisano2012}
A.~Pisano and Y.~Orlov, ``Boundary second-order sliding-mode control of an
  uncertain heat process with unbounded matched perturbation,''
  \emph{Automatica}, vol.~48, no.~8, pp. 1768--1775, 2012.

\bibitem{wang2015}
J.-M. Wang, J.-J. Liu, B.~Ren, and J.~Chen, ``Sliding mode control to
  stabilization of cascaded heat {PDE–ODE} systems subject to boundary
  control matched disturbance,'' \emph{Automatica}, vol.~52, pp. 23--34, 2015.

\bibitem{brezis1973operateurs}
H.~Brezis, \emph{Op{\'e}rateurs maximaux monotones et semi-groupes de
  contractions dans les espaces de Hilbert}.\hskip 1em plus 0.5em minus
  0.4em\relax Elsevier, 1973.

\bibitem{fenza2025finite}
K.~Fenza, M.~Labbadi, and M.~Ouzahra, ``Finite-time stabilization of a class of
  nonlinear systems in hilbert space,'' \emph{in: Proceedings of the 64th IEEE
  Conference on Decision and Control}, 2025.

\bibitem{pazy2012semigroups}
A.~Pazy, \emph{Semigroups of linear operators and applications to partial
  differential equations}.\hskip 1em plus 0.5em minus 0.4em\relax Springer
  Science, 1983.

\bibitem{engel2000one}
K.-J. Engel, R.~Nagel, and S.~Brendle, \emph{One-parameter semigroups for
  linear evolution equations}.\hskip 1em plus 0.5em minus 0.4em\relax Springer,
  2000, vol. 194.

\bibitem{polyakov2021input}
A.~Polyakov, ``Input-to-state stability of homogeneous infinite dimensional
  systems with locally lipschitz nonlinearities,'' \emph{Automatica}, vol. 129,
  p. 109615, 2021.

\bibitem{polyakov2024finite}
A.~Polyakov and Y.~Orlov, ``Finite/fixed-time homogeneous stabilization of
  infinite dimensional systems,'' \emph{IEEE Transactions on Automatic
  Control}, 2024.

\bibitem{bhat2000finite}
S.~Bhat and D.~Bernstein, ``Finite-time stability of continuous autonomous
  systems,'' \emph{SIAM Journal on Control and Optimization}, vol.~38, pp.
  751--766, 2000.

\bibitem{bhat2005geometric}
------, ``Geometric homogeneity with applications to finite-time stability,''
  \emph{Mathematics of Control, Signals, and Systems}, vol.~17, pp. 101--127,
  2005.

\bibitem{polyakov2011nonlinear}
A.~Polyakov, ``Nonlinear feedback design for fixed-time stabilization of linear
  control systems,'' \emph{IEEE transactions on Automatic Control}, vol.~57,
  no.~8, pp. 2106--2110, 2011.

\bibitem{jimenez2020lyapunov}
E.~Jim{\'e}nez-Rodr{\'\i}guez, A.~J. Mu{\~n}oz-V{\'a}zquez, J.~D.
  S{\'a}nchez-Torres, M.~Defoort, and A.~G. Loukianov, ``A lyapunov-like
  characterization of predefined-time stability,'' \emph{IEEE Transactions on
  Automatic Control}, vol.~65, no.~11, pp. 4922--4927, 2020.

\bibitem{curtain2016stabilization}
R.~Curtain and H.~Zwart, ``Stabilization of collocated systems by nonlinear
  boundary control,'' \emph{Systems \& control letters}, vol.~96, pp. 11--14,
  2016.

\bibitem{curtain2020introduction}
------, \emph{Introduction to infinite-dimensional systems theory: a
  state-space approach}.\hskip 1em plus 0.5em minus 0.4em\relax Springer
  Nature, 2020, vol.~71.

\bibitem{clarke2013functional}
F.~Clarke, \emph{Functional analysis, calculus of variations and optimal
  control}.\hskip 1em plus 0.5em minus 0.4em\relax Springer, 2013, vol. 264.

\bibitem{parsegov2012nonlinear}
S.~Parsegov, A.~Polyakov, and P.~Shcherbakov, ``Nonlinear fixed-time control
  protocol for uniform allocation of agents on a segment,'' in \emph{2012 IEEE
  51st IEEE conference on decision and control (CDC)}.\hskip 1em plus 0.5em
  minus 0.4em\relax IEEE, 2012, pp. 7732--7737.

\end{thebibliography}
\begin{IEEEbiography}[{\includegraphics[width=1in,height=1.25in,clip,keepaspectratio]{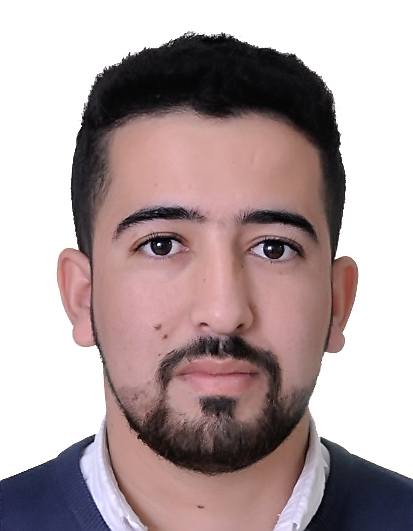}}]{Kamal Fenza}
 is a PhD student in the MASI Laboratory, Department of Mathematics and Informatics, ENS, University of Sidi Mohamed Ben Abdellah, Morocco. He received his Master degree in Applied Mathematics in 2022 from the École Normale Supérieure (ENS), University of Sidi Mohamed Ben Abdellah, Fez. His research area focused  on the finite and fixed-time stability of distributed  nonlinear systems, control theory, and their applications.

\end{IEEEbiography}

\begin{IEEEbiography}[{\includegraphics[width=1in,height=1.25in,clip,keepaspectratio]{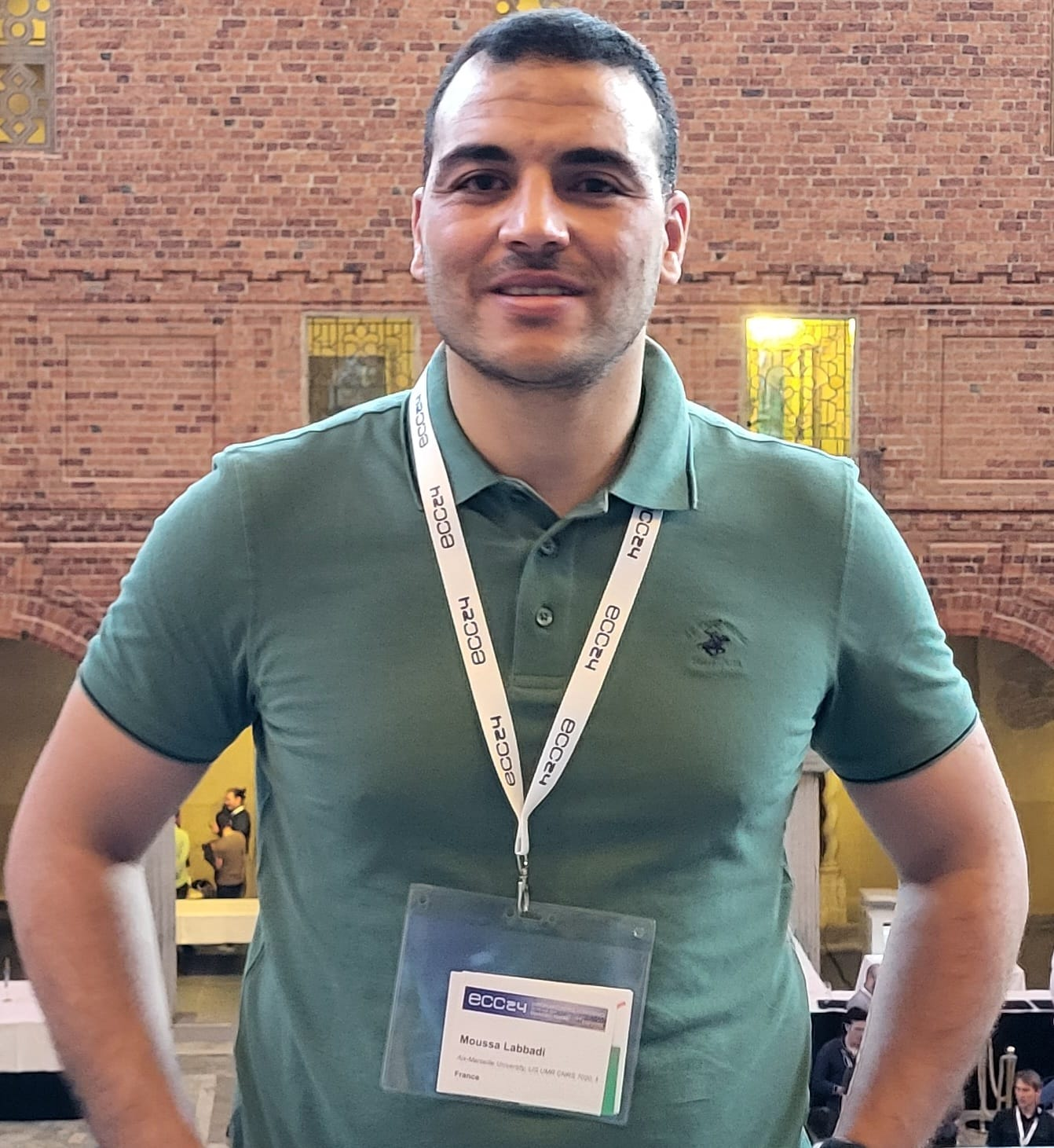}}]{Moussa Labbadi}
(Member, IEEE) received the B.S. degree in Mechatronics Engineering from UH1 in 2015, the M.S. degree in Mechatronics Engineering from UAE in 2017, and the Ph.D. degree in Automatic Control from EMI, UM5, Rabat, Morocco, in September 2020. From 2020 to 2021, he served as a Researcher at the MAScIR Foundation, Morocco. Subsequently, from 2021 to 2022, he held the position of ATER with the Department of Automatic Control, INSA, LAMIH, and UPHF. Between 2022 and 2023, he worked as a Postdoctoral Researcher at INPG, GIPSA-Lab, and UGA. He is currently an associate professor at Aix-Marseille University (AMU) and a member of the Laboratory of Informatics and Systems (LIS).
His research interests span variable structure control, sliding mode control, observation, nonlinear system stability, control theory, and their applications.
Dr. Labbadi is an active member of several IEEE Technical Committees.

\end{IEEEbiography}

\begin{IEEEbiography}[{\includegraphics[width=1in,height=1.25in,clip,keepaspectratio]{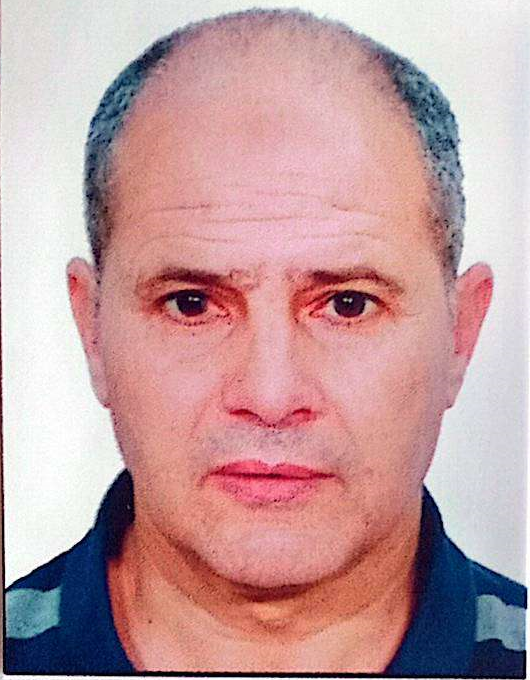}}]{Mohamed Ouzahra}
 received his Ph.D. in Control Theory of Distributed Systems from Moulay Ismail University, Morocco, in 2004. He is currently a Full Professor of Mathematics at the Graduate Normal School (ENS), University of Sidi Mohamed Ben Abdellah (USMBA),Fez, Morocco.\\
He is a member of the Laboratory of Mathematics and Applications to Engineering Sciences.
His research interests lie in the field of control theory, with a particular emphasis on the analysis and control of distributed parameter systems. His current work focuses on three main directions: Feedback stabilization of bilinear systems, multiplicative controllability of partial
differential equations (PDEs) and optimal control problems for bilinear systems.

\end{IEEEbiography}

\end{document}